\documentclass[conference]{IEEEtran}
\IEEEoverridecommandlockouts
\UseRawInputEncoding
\usepackage{graphicx}
\usepackage{amsmath}
\usepackage{amsfonts}
\usepackage{mathrsfs}
\usepackage[OT1]{fontenc} 
\usepackage{enumerate}
\usepackage{amsthm}
\usepackage{tabularx}
\usepackage{arydshln}

\usepackage[colorlinks=true,linkcolor=blue,citecolor=blue]{hyperref}
\usepackage{mwe}
\usepackage{subfig}
\usepackage{float}
\usepackage{graphicx}
\usepackage{textcomp}
\usepackage{xcolor}
\usepackage{footnote}
\usepackage{makecell}
\usepackage{graphicx}
\usepackage{amsmath}
\usepackage{amsfonts}
\usepackage{mathrsfs}
\usepackage[OT1]{fontenc} 
\usepackage{enumerate}
\usepackage{amsthm}
\usepackage{tabularx}
\usepackage{arydshln}
\hypersetup{hidelinks}

\usepackage{subfig}
\usepackage{textcomp}
\usepackage{xcolor}
\usepackage{footnote}
\usepackage{makecell}
\usepackage[draft]{todonotes}

\usepackage{algorithm}
\usepackage{algpseudocode}
\usepackage[inline]{enumitem}
\usepackage{mathtools}

\definecolor{bleudefrance}{rgb}{0.19, 0.55, 0.91}
\definecolor{chromeyellow}{rgb}{1.0, 0.65, 0.0}
\definecolor{asparagus}{rgb}{0.53, 0.66, 0.42}

\begin{document}
\title{Framework for Highway Traffic Profiling using Connected Vehicle Data\\
\thanks{}}


\author{
\IEEEauthorblockN{ Zijia Zhong, Liuhui Zhao, Branislav Dimitrijevic, 
Dejan Besenski, Joyoung Lee} 
\IEEEauthorblockA{John A. Reif, Jr. Department of Civil and Environmental Engineering\\
\textit{New Jersey Institute of Technology}\\
Newark, NJ, USA \\
\{zijia.zhong, liuhui.zhao, bxd1947, besenski, jo.y.lee\}@njit.edu}
}

\maketitle
\begin{abstract}
The connected vehicle (CV) data could potentially revolutionize the traffic monitoring landscape as a new source of CV data that are collected exclusively from original equipment manufactures (OEMs) have emerged in the commercial market in recent years.  Compared to existing CV data that are used by agencies, the new-generation of CV data have certain advantages including nearly ubiquitous coverage, high temporal resolution, high spatial accuracy, and enriched vehicle telematics data (e.g., hard braking events). This paper proposed a traffic profiling framework that target vehicle-level performance indexes across mobility, safety, riding comfort, traffic flow stability, and fuel consumption. The proof-of-concept study of a major interstate highway (i.e., I-280 NJ), using the CV data, illustrates the feasibility of going beyond traditional aggregated traffic metrics. Lastly,  potential applications for either historical analysis and even near real-time monitoring are discussed. The proposed framework can be easily scaled and is particularly valuable for agencies that wish to systemically monitoring regional or statewide roadways without substantial investment on  infrastructure-based sensing (and the associated on-going maintenance costs). 
 

\end{abstract}



\section{Introduction}

Connected vehicle (CV) data have been widely used for numerous transportation applications with proven track records. With data being generated  at a scale that ones never seen before and the proliferation of mobile sensors (e.g., smart handheld devices), the availability of CV data have been rapidly expanding. The landscape of CV data have transitioned from floating car run with only a handful of vehicle to commercial data vendors (e.g., INRIX, Wejo) that aggregated millions of equipped vehicle data points from commercial fleet operation to passenger vehicles. For instance, The INRIX Trip Analytics is one of such datasets. It reportedly covers 10\% of all vehicle trips in the United States \cite{NREL2020Examines}. However, note that the 10\% penetration rate are not uniformly distributed in the spatial and temporal dimensions, which means coverage could be sparse at certain remote areas or during night-time hours. Data collected from smartphone with inherent low-power consumption profile tends to be more sparse, compared to on-board units.

Recently, a new form of commercialized CV data (e.g., Wejo, Otonomo, High-Mobility) that are collected exclusively from OEMs (Original Equipment Manufacturer) vehicles has emerged.  These types of data, while varies across difference OEMs, are collected via the OEM's telematics system with the built-in wireless communication capability in the late model cars. The granularity of such commercialized CV data is one of the major advantages compared to the ad hoc CV data or CV data fused with hand-held devices. This presents an immense potential for transportation planning, operation \& maintenance, and incident management. Besides high-resolution waypoint data,  These emerging data sets also provide vehicle telemetry data (e.g., fuel level, ABS engagement, hard braking, windshield wiper activation, etc.). Based on Wejo's (a leading data provider) internal estimation, it receives information from 1 in 20 vehicle in U.S. (and 1 in 50 vehicles in Europe). Otonomos is reportedly has more than 4 billions data points from over 40 million licensed vehicles ingested into its platform \cite{Otonomo}.

In the advent of commercial large-scale CV data, this paper proposes an traffic profiling framework that ingest massive amount of commercial CV data with waypoints and telemetry information.  This illustrates the observability and benefits for agencies and practitioners to utilize the CV data at scale in a cost effective way.

This paper proposes a roadway traffic profiling framework that utilize large-scale commercial CV data. The highlights of this study include: 
\begin{enumerate*}[label=\roman*)]
\item a monitoring framework that ingest large-scale CV data \item the traffic performance indexes that reaches a new-level of granularity beyond traditional traffic performance measures, and  \item the proof-of-concept case study that inspires future utilization of the CV data.

\end{enumerate*}


\section{Literature Review}
\subsection {Connected Vehicle Data}

CV data have been used for providing observability of our transportation systems. It was found that CV data with only 1.5-2\% penetration is sufficient to infer the vehicle-mile travelled (VMT) for the State of Maryland \cite{fan2019using}. CV date have also been used to develop GPS-based automated traffic signal performance measures (ATSPM) that have expanded coverage and scalability. A quick diagnosis of the immediate signal performance issues can be identified with as low as 0.04\%\cite{waddell2020utilizing}. The queue propagation around freeway bottleneck and congestion on arterial based on the percentage of slow-moving vehicles can be identified using CV data \cite{khadka2022developing}.

In the State of Indiana, Wejo data was compared against 24 traffic count locations on Indiana roadways and 4.3\% MPR  with standard deviation (STDEV) 1.0\% and 5.0\% MPR (STDEV 1.36\%) \cite{hunter2021estimation} for interstates and non-interstate roads, respectively.
In New Jersey roadway, our study \cite{dimitrijevic2022assessing} showed that the observed market penetration rate for the Wejo Movement data for the interstate highways, non-interstate expressway, major arterial, and minor arterial are  2.55\% (std. dev. 0.76 \%),  2.31\% (std. dev. 1.07 \%),  3.25\% (standard deviation 1.48 \%), and 4.39\% (standard deviation 2.65\%), respectively. Studies also reported extreme MPRs in smaller and more specific regions. In the Dallas-Fort Worth area in Texas, the Wejo data presented 10\%-15\% of all moving vehicles \cite{khadka2022developing}. Approximately 10\% penetration was observed the area around an OEM facility \cite{hunter2021estimation}. On the other end of the spectrum, a study that used Wejo trajectory data (collected in 2019) for estimating boarder crossing times reported that around 30\% of the test hours with no Wejo samples at the Paso del Norte port of entry in El Paso, TX. It is worth noting that this study also confirmed the strong correlation between the Wejo travel time (when there was samples collected)  and the Bluetooth travel time within the boarder crossing area.\cite{li2021exploring}

\subsection{CV Applications}
The Wejo hard-braking event data was used as a safety surrogate measure for a before-and-after study for a left-turn phase (located in Indiana) that was converted from protected-permitted to protected left turn. The hard-braking event is recorded when a on-board accelerometer measures a acceleration that is greater than $2.638 m/s^{2}$. The hard-braking counts and percentage of vehicles turning left were compared, and the study found a 14\% reduction on hard-braking events after the conversion \cite{saldivar2021using}. In another study, the Wejo hard-breaking event data was explored for agency-wide screening of intersections, which is typically rely on 3-5 years of crash data. Strong correlation between rear-end crashes and hard-braking events that occurred past 400 ft. of the stop bar was revealed. In addition, the study concluded that the hard-braking data enable agencies to address emerging problem with quicker than typical practices using multi year crash data.

Saldivar-Carranza et al. \cite{saldivar2021longitudinal} utilized Wejo trajectory data to assess the traffic impact to local arterials, especially for the unofficial detour that enabled by personal navigation devices, such as Google Map Navigation.   During the 11-week period, the weekly afternoon peak period volumes, split failures, downstream blockage, arrivals on green, and travel time were monitored. \cite{saldivar2021diverging} is another similar use case of the near ubiquity Wejo data for locations with little or non-existent sensing infrastructure.

Khadka et al. \cite{khadka2022developing} used Wejo trajectory data to directly measure queue length and its propagation on freeway bottlenecks. A interstate segment in Arlington, TX was discretized into 0.5-mi segments and local empirical speed threshold (i.e., 45 mph) was used. Though the sample trajectory are only a subset of the overall traffic stream, high correlation between travel time and slow trajectories was observed.


\subsection{Profiling Metrics}
In terms of roadway performance measures, they can be grouped into three categories: mobility, safety, environmental impact (e.g., emission and fuel consumption). Travel time, throughput, and speed are the main metrics within the mobility category. In the past decade, the reliability dimension was integrate into the mobility metrics, such as planning time index, travel time index. A new series of highway traffic profiling metrics have become available with increase data complexity and direct measure at individual vehicle level can be achieved.

Total ADT if found to significantly impact crash frequency: for every 1\% change in ADT, the crash frequency is expected to increase by 0.43\% \cite{schneider2010effects}. Speed is also a significant contributing factor for roadway safety. In 2016, 18\% of drivers involved in fatal crash in US were speeding at the time of crashing \cite{nhtsa2016traffic} (such number increased to 29\% in work zones \cite{fhwa2016workzone}).
It was found that 12 mph above the average speed is the threshold where crash risk was minimized. Larger speed variations posed a less risk than speeds that are significantly lowers than the average speed.
The speed variation, measured as the difference between the 85th percentile speed and average speed was found to be one of the primary factor affecting crash rate. 
It was shown that 1 mph increase in the standard deviation of speed would lead to a 27.8\% increase in the total number of crashes \cite{day2019evaluation}. In one study, large speed variation was observed 6-12 minutes prior to crash could increase the possibility of severe crashes \cite{xu2016evaluation}. Another study found 1\% increase in speed variation was related to 0.74\% higher crash frequency \cite{wang2018speed}.

Jerk has been also used as an indicator for detecting safety critical event\cite{bagdadi2013development}, traffic conflicts\cite{zaki2014use}, driving aggressiveness\cite{feng2017can}, and change of driving decision \cite{liu2015role}. Longitudinal jerk (change of acceleration) is mathematically defined as the derivative of acceleration. Vehicle jerk is related to a driver's physiological feeling of ride comfort. For discomfort, three important aspects were identified, which are the peak of jerk, the frequency of jerk, and the duration of exposure to jerk \cite{huang2004fundamental}. For instance, the International Organization for Standardization (ISO) set the threshold for the negative jerks of the vehicle as $-2.5m/{s^3}$ during automatic braking for adaptive cruise control (ACC) \cite{iso2010intelligent}. A large jerk when accompany with a relatively large acceleration should be paid attention for improving ride comfort \cite{huang2004fundamental}.
A study conducted by Fend et al. found that aggressive drivers identified by three metrics: speeding, tailgating and with crash or near-crashes incidents are correlated with large positive jerk (greater than $1.07m/{s^3}$) and large negative jerk (smaller than $-1.47/{s^3}$) \cite{feng2017can}.




\section{Monitoring Framework}

\subsection{Data}
The CV data used in the study (acquired from Wejo Group Ltd.) is exclusively crowd-sourced from OEMs. It provides vehicle waypoints (latitude and longitude) and timetimestamp, instantaneous speed, vehicle heading along with other metrics. The majority of the ping interval for consecutive waypoints for a vehicle is found to be 3 seconds. The spatial resolution of the data is with 6-digit decimal point, which is approximately 3-meters resolution (lane-level resolution). 
The volume of the data, given its granularity, sample size, and coverage, poses a challenge in terms of data storage, processing, visualization, and analytics. For instance, the CV data set for New Jersey in June 2021 contains to 8.17 billion data points from 22.18 million journeys. Additionally, we also found the equipped vehicles were evenly distributed throughout the traffic stream. For interstate highways during the day time, most of the headway between consecutive vehicles registered at a weight-in-motion count stations are less than 40 seconds.

\subsection{Profiling Metrics}
To generate an entire picture of traffic conditions, different categories of metrics could be established from the individual vehicle movement data. To reflect different aspects of traffic conditions, we calculate various measurement from movement data and categorize them into different groups, including: 
\begin{enumerate}[label=\roman*)]
\item Overall mobility condition per segment per time interval, including number of vehicles (i.e., traffic volume) $n_{sk}$, average traffic speed $mu_{sk}$, and number of waypoints $m_{sk}$, where $s \in [0,...,S]$ is the index of a 0.5-mile segment, and $k \in [0,...,K]$ is the study time interval, which is 30-min in this case study. Although the CV only a subset of the overall traffic stream, it can reveal the deviation for the historical traffic pattern.
\item Safety index, which is a combination from speed coefficient of variation, percentage of speed drop from speed limit, and the average absolute heading change along each segment.
For each 0.5-mile segment, we define $\sigma_{sk}$ as the standard deviation, $v^d_s$ as the posted speed limit. The speed coefficient of variation $v^c_{sk}$ is defined as the ratio of standard deviation and average speed, i.e.,. ${\sigma_{sk}}/{\mu_{sk}}$. Let $v^r_{sk}$ represents the percentage of speed drop from speed limit, and define $v^r_{sk}$ as $({\mu_{sk} - v^d_{s}})/{v^d_{s}}$. Let $i$ be the index of vehicle available during time interval $k$ along segment $s$, $t \in [0,...,T]$ be the index of time point in the study interval $k$, $h_i^t$ be the heading of vehicle $i$ at timestamp $t$, and $h^c_{sk}$ be the average absolute heading changes along the 0.5-mile segment. We define $h^c_{sk}$ as $\big({\sum_i \frac{\sum_t \left | h_i^t -h_i^{t-1}\right |}{360} } \big)/ {n_{sk}}$. To integrate the three measurements, the average absolute heading change is normalized to a ratio over 360 degree.

Then, the safety index for each segment during the 30-min interval, denoted as $I_{sk}$, is defined as weighted sum of these three measurements, as in Eq. \ref{eq:safety}, where $w_{vc}$, $w_{vr}$, $w_{hc}$ are the weights for $v^c_{sk}$, $v^r_{sk}$, and $h^c_{sk}$, respectively.

\begin{equation}
\label{eq:safety}
I_{sk} = w_{vc} \cdot v^c_{sk} + w_{vr} \cdot v^r_{sk} + w_{hc} \cdot h^c_{sk}
\end{equation}

\item Comfort index, which consists of percentage of brakes and percentage of high jerks during the study period. As reported by the literature, jerk could influences the physiological feeling of the drivers, the human factor aspect of traffic safety. Let $p^b_{sk}$ be the percentage of brakes and $p^j_{sk}$ be the percentage of high jerks over all the waypoints during study time interval $k$ along the 0.5-mile segment $s$. The comfort index $C_{sk}$ is defined as the weighted sum of $p^a_{sk}$ and $p^j_{sk}$, as shown in Eq. \ref{eq:comfort}.

\begin{equation}
\label{eq:comfort}
C_{sk} = w_{pb} \cdot p^b_{sk} + w_{pj} \cdot p^j_{sk}
\end{equation}

\item Stability index, which is represented by number of large brakes and large accelerations. Let $n^a_{sk}$ be the number of large accelerations and $n^b_{sk}$ be the number of large brakes during study time interval $k$ along the 0.5-mile segment $s$. For the analysis purpose, large brakes and large accelerations can be set using the thresholds for hard brake/acceleration or uncomfortable brake/acceleration values. The stability index $Y_{sk}$ is then defined as the weighted sum of $n^a_{sk}$ and $n^b_{sk}$, as shown in Eq. \ref{eq:stability}.

\begin{equation}
\label{eq:stability}
Y_{sk} = w_{na} \cdot n^a_{sk} + w_{nb} \cdot n^b_{sk}
\end{equation}

\item Fuel consumption, which is estimated by individual vehicle acceleration and speed information. Most vehicle trajectory-based module are suitable for the monitoring application. In this study a polynomial fuel consumption model (Eq. \ref{eq:fuel}) that produces instantaneous, second-by-second fuel consumption were adopted \cite{kamal2012model}. 

\begin{subequations}
\label{eq:fuel}
\begin{gather} 
f_v = f_{cruise} + f_{accel}\\
f_{cruise} = b_0 + b_1 v + b_2 v^2 + b_3 v^3  \\
f_{accel}=a(c_0 + c_2 v + c_2 v^2)
\end{gather}
\end{subequations}
where $f_{cruise}$ is fuel consumption per seconds at a steady speed of $v$, $f_{accel}$ is additional consumption due to acceleration $a$ at speed $v$, and all other constants were empirical parameters for engine fuel consumption.

\end{enumerate}


\section{Case Study}
\begin{figure}[!ht]
    \centering
    \includegraphics[width = 0.45\textwidth]{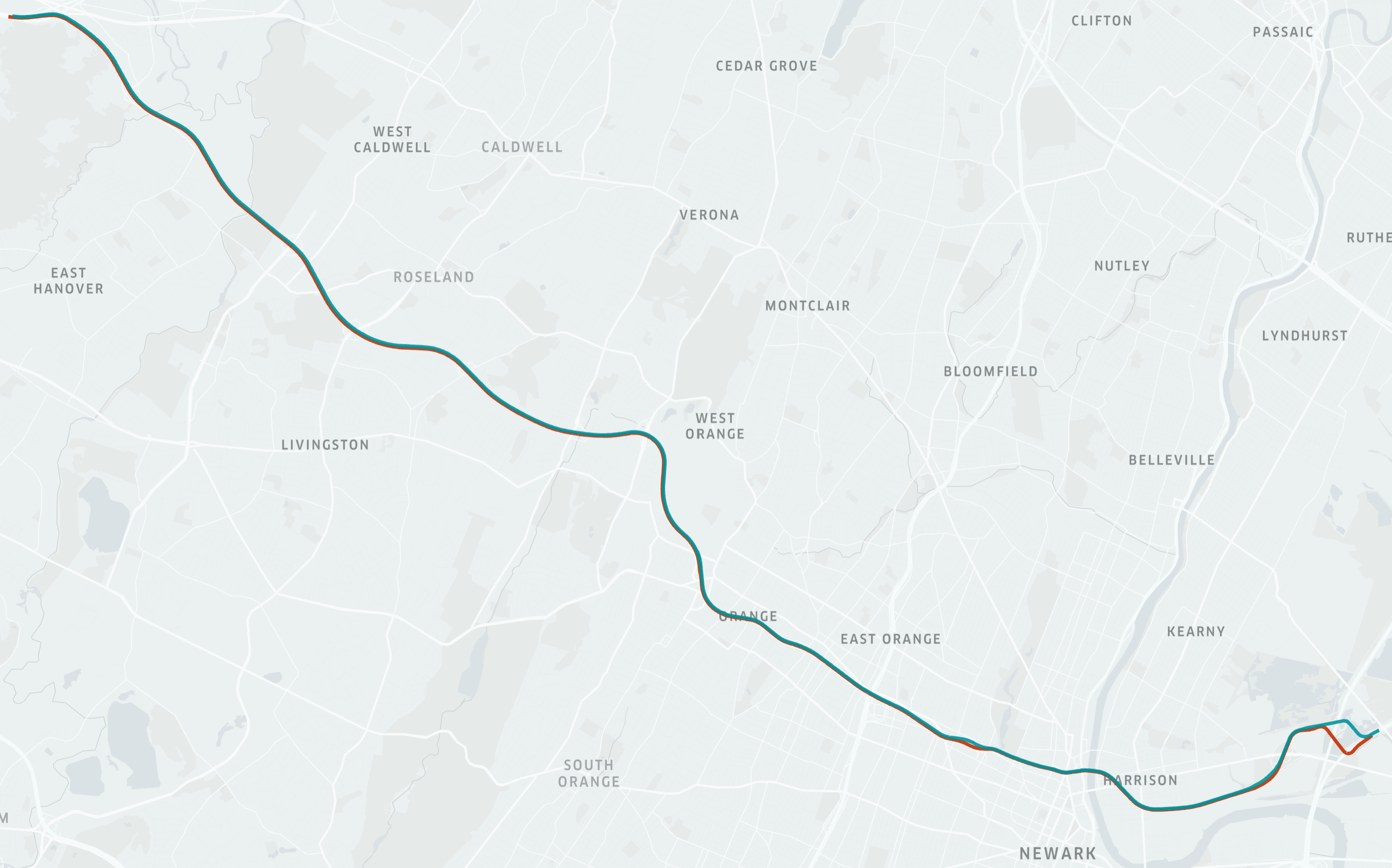}
    \caption{I-280 in New Jersey.}
    \label{fig:I280}
\end{figure}

To illustrate highway traffic profiling with high-resolution vehicle movement data, we selected interstate highway I-280 in New Jersey as a case study. I-280 is a east-west major commuter corridor that spans 17.85 miles (28.73 km), as shown in Fig. \ref{fig:I280}. The last stretch to the east of the roadway runs through urban area while the western part crosses suburban mountainous area. The east end of the roadway connects New Jersey Turnpike featuring a toll plaza, providing access towards tunnels to New York City. 

The roadway is broken down to 0.5-mile segment for each direction for analysis purpose. The metrics are calculated for each segment such that traffic profiles are established at every segment during different time periods. Two day data (2021-05-23 and 2021-06-06) were processed for comparison purpose, with several crashes on record as shown in Table \ref{tab:crash}. According to NJDOT crash record, a fatal crash happened on I-280 eastbound between exit 8 and exit 10 (milepost 9.9) at 1:15PM, leading to all lanes closed about an hour later for crash mitigation and investigation, and two injury related crashes happened on 2021-05-23 during morning hours.

\begin{table}[!ht]
    \renewcommand\arraystretch{1.2}
    \caption{crash records during study period}
    \centering
    \begin{tabular}{|c|c|c|c|}\hline
        Date \& Time & Direction & Milepost & Severity \\\hline
        2021-06-06 13:15 & EB & 9.9 & Fatal Injury\\
        2021-05-23 02:05 & EB & 14.2 & Possible Injury\\
        2021-05-23 09:28 & WB & 4.6 & Suspected Minor Injury\\\hline
    \end{tabular}
    \label{tab:crash}
\end{table}

\subsection{Traffic Overview}
We take 30-min as the study interval for illustration purpose, which could be adjusted to different interval as needed (e.g., 1-min, 5-min, or 1-hr). Thus, a set of metrics, including number of vehicles, number of waypoints, average speed, etc., are aggregated every 30-min for each 0.5-mile segment.

\subsubsection{Number of vehicles} 
\begin{figure}[h]
\begin{minipage}[h]{0.45\textwidth}
\centering
\subfloat[eastbound]{\includegraphics[scale=0.23]{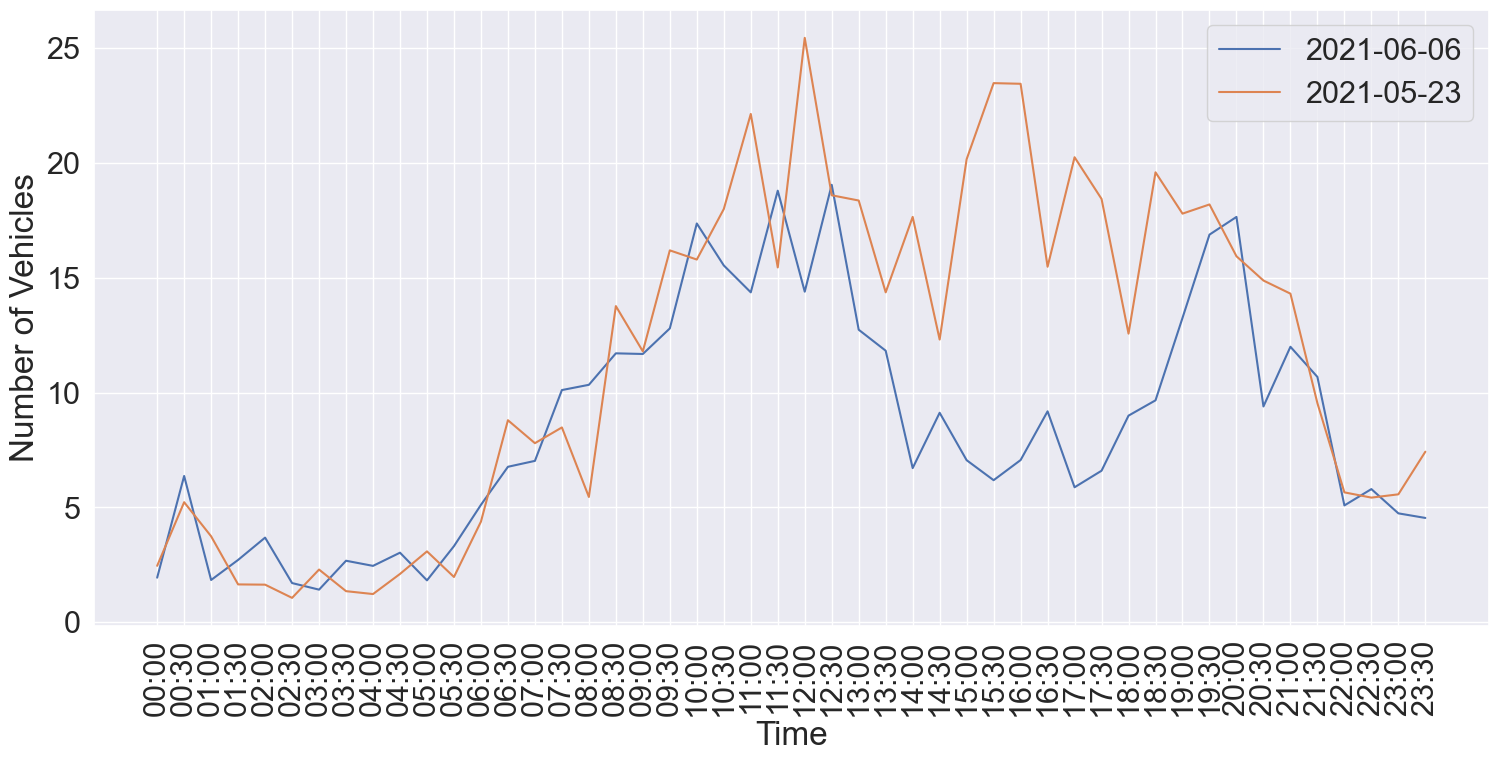}}
\end{minipage} \\
\begin{minipage}[h]{.45\textwidth}
\centering
\subfloat[westbound]{\includegraphics[scale=0.23]{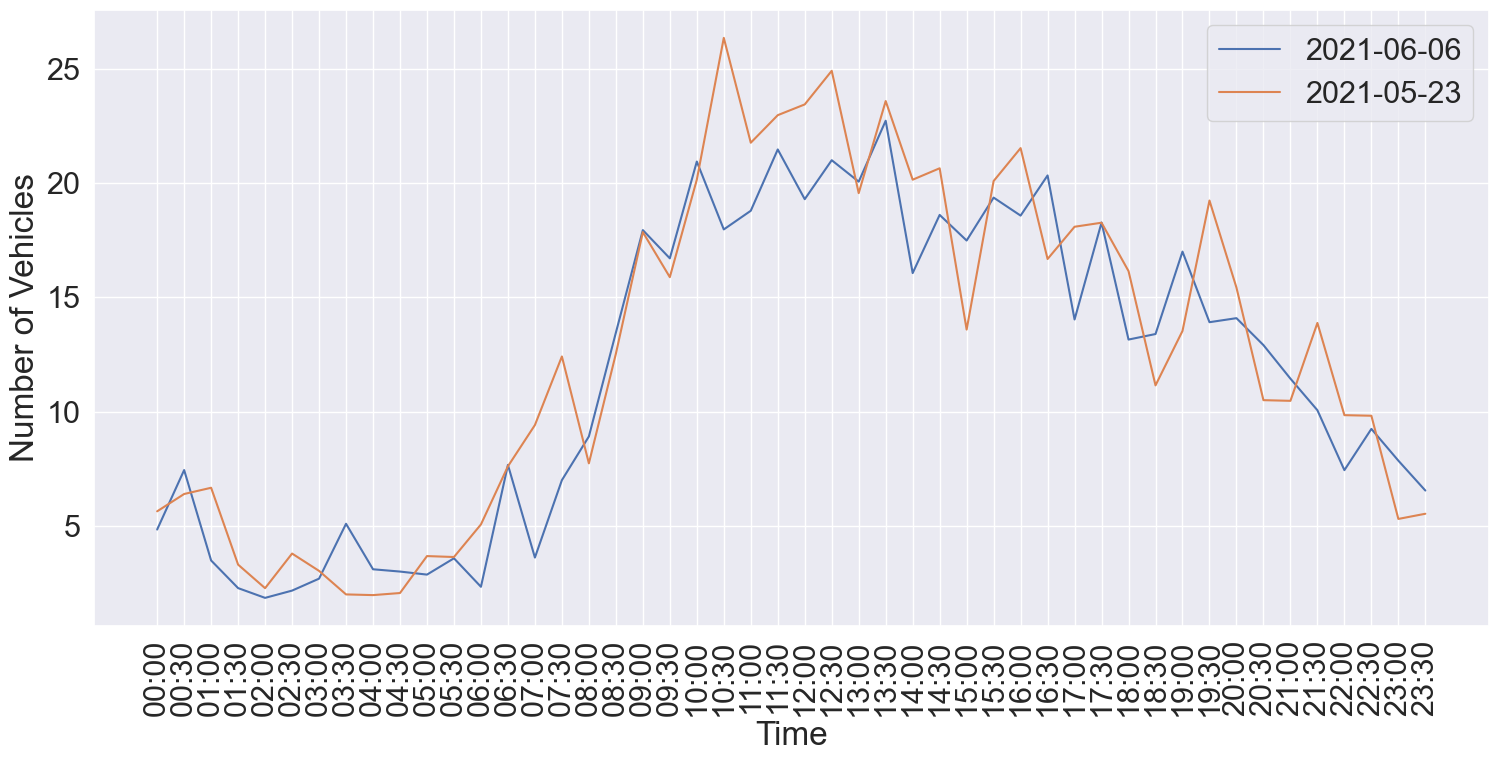}}
\end{minipage}
\caption{Number of vehicles (Time-of-day variation) }
\label{fig:nvehs_time}
\end{figure}

the first measurement obtained from the vehicle movement data is the number of vehicles per 30-min per segment. Fig. \ref{fig:nvehs_time} represents the average number of vehicles for each 30-min interval. The upper plot shows the eastbound traffic, with a big gap in the afternoon hours due to the fatal crash and lane closure event on 2021-06-06. On the contrary, without major disruption, westbound traffic shows similar pattern on two days.

\subsubsection{Average speed} the average speed is calculated by averaging the speeds from all the waypoints during 30-min interval for each 0.5-mile segment. Overall, we observe a lower traffic speed towards the eastern part (i.e., urban area) of I-280 with intense on and off ramps for both eastbound and westbound traffic. It is also reasonable that the last segment with toll plaza experience consistently low speed.

\begin{figure}[H]
\begin{minipage}[h]{0.45\textwidth}
\centering
\subfloat[eastbound]{\includegraphics[scale=0.14]{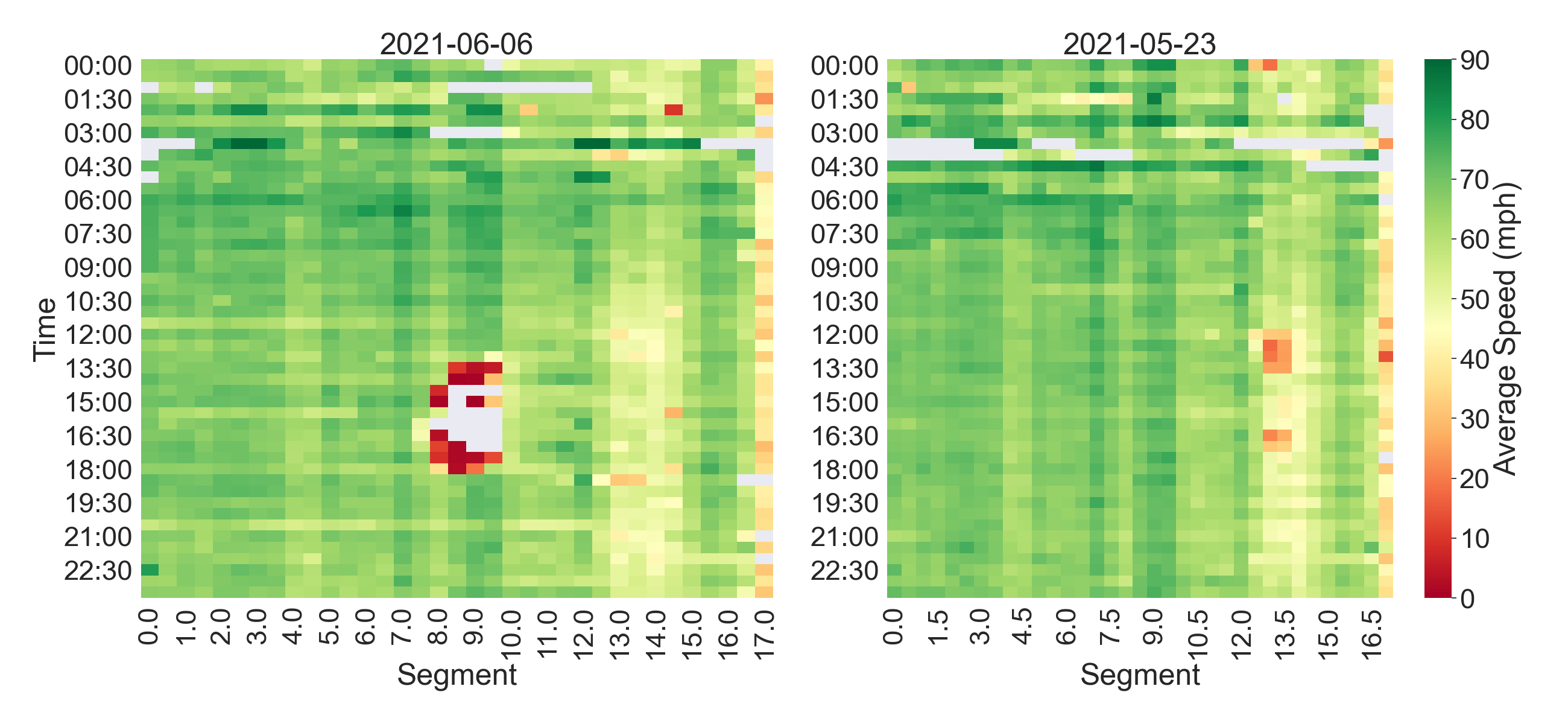}}
\end{minipage} \\
\begin{minipage}[h]{.45\textwidth}
\centering
\subfloat[westbound]{\includegraphics[scale=0.14]{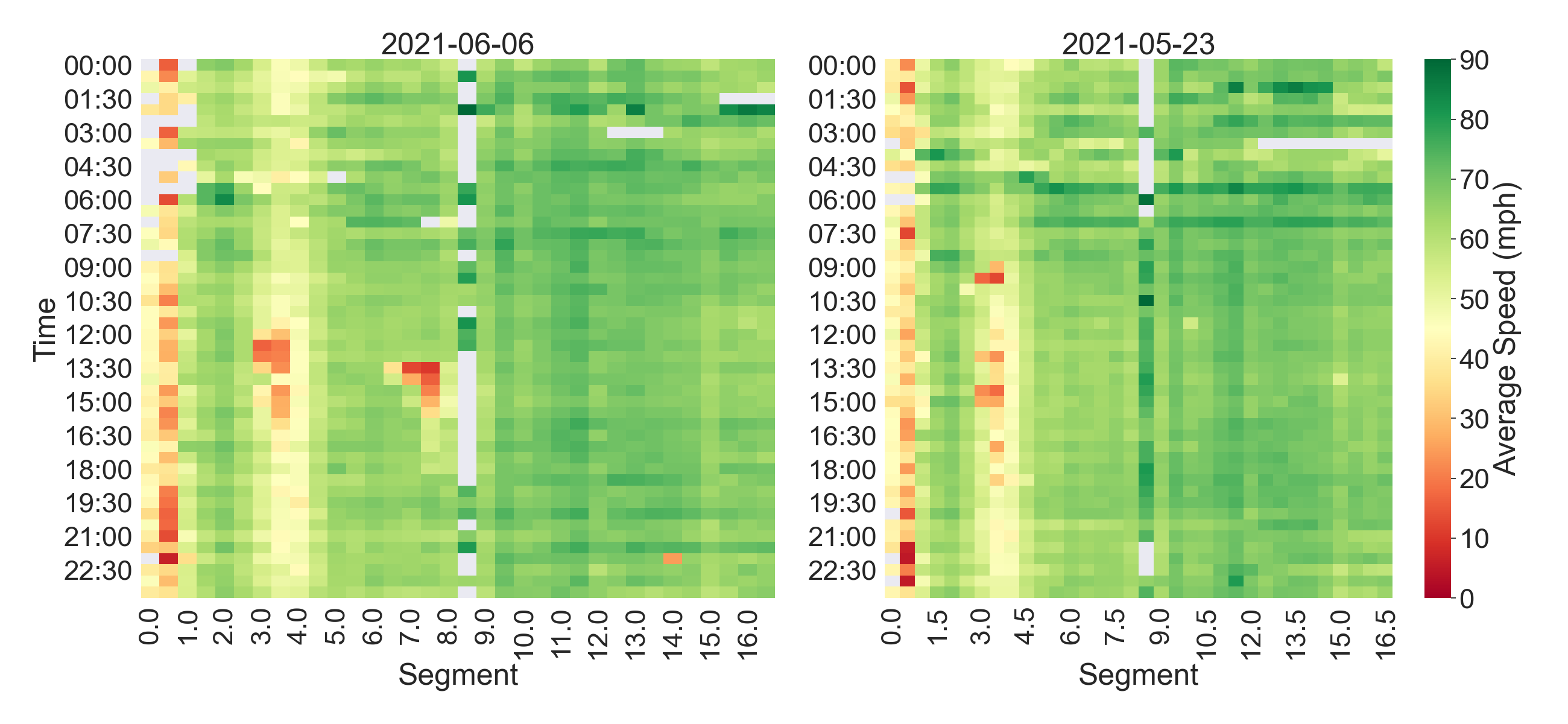}}
\end{minipage}
\caption{Average Traffic Speed}
\label{fig:avg_spd_heat}
\end{figure}

Specifically, when the fatal crash occurred on eastbound segment, significant congestion is observed on the upstream segment (milepost 8.5 to 10.0) with traffic speed dropping below 10mph, and up to 2-mile upstream traffic was affected by this event, as shown in Fig. \ref{fig:avg_spd_heat}. It is interesting to observe that the fatal crash in the eastbound segment also led to speed reduction and traffic disruption on the corresponding westbound segment, as shown in Fig. \ref{fig:avg_spd_heat}.

\subsubsection{Number of waypoints} 
\begin{figure}[h]
\begin{minipage}[h]{0.45\textwidth}
\centering
\subfloat[eastbound]{\includegraphics[scale=0.14]{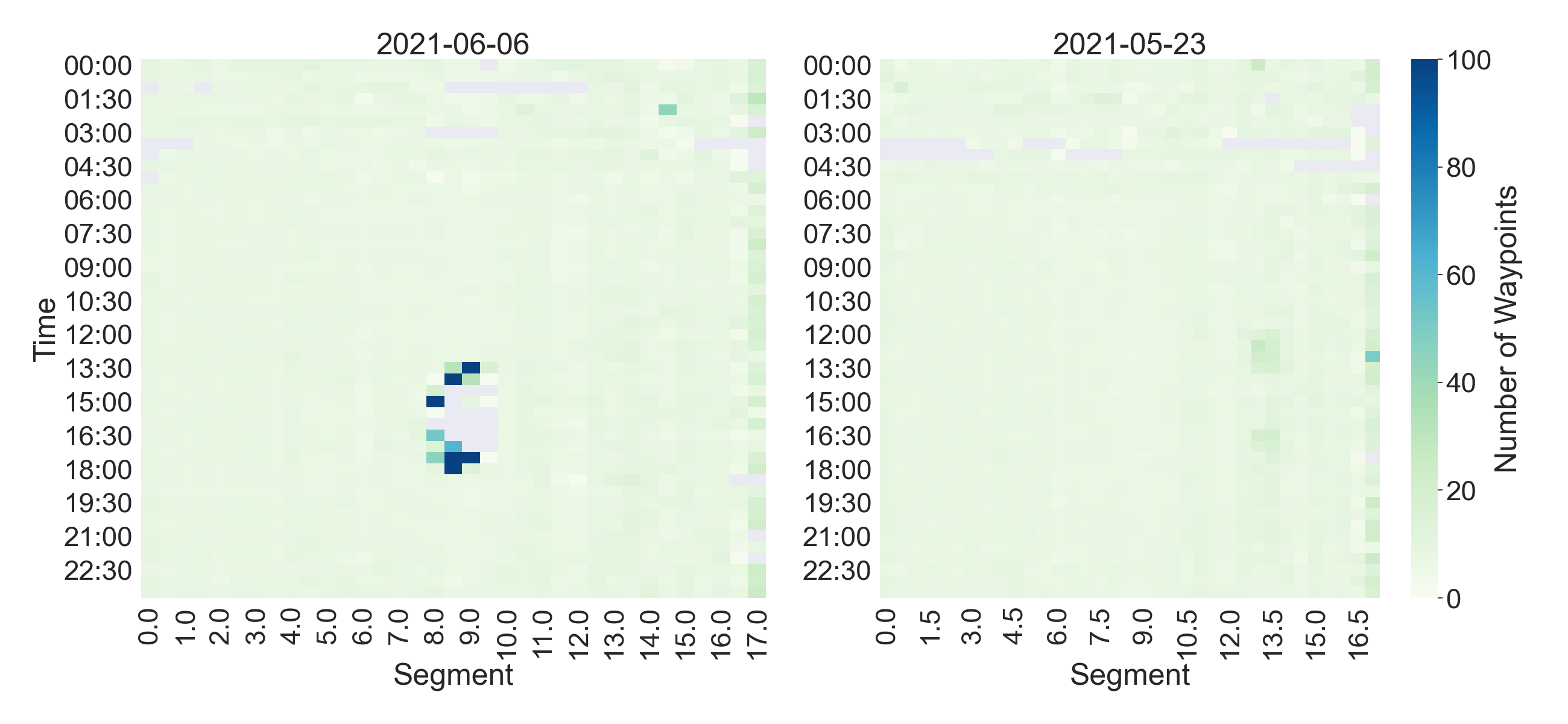}}
\end{minipage} \\
\begin{minipage}[h]{.45\textwidth}
\centering
\subfloat[westbound]{\includegraphics[scale=0.14]{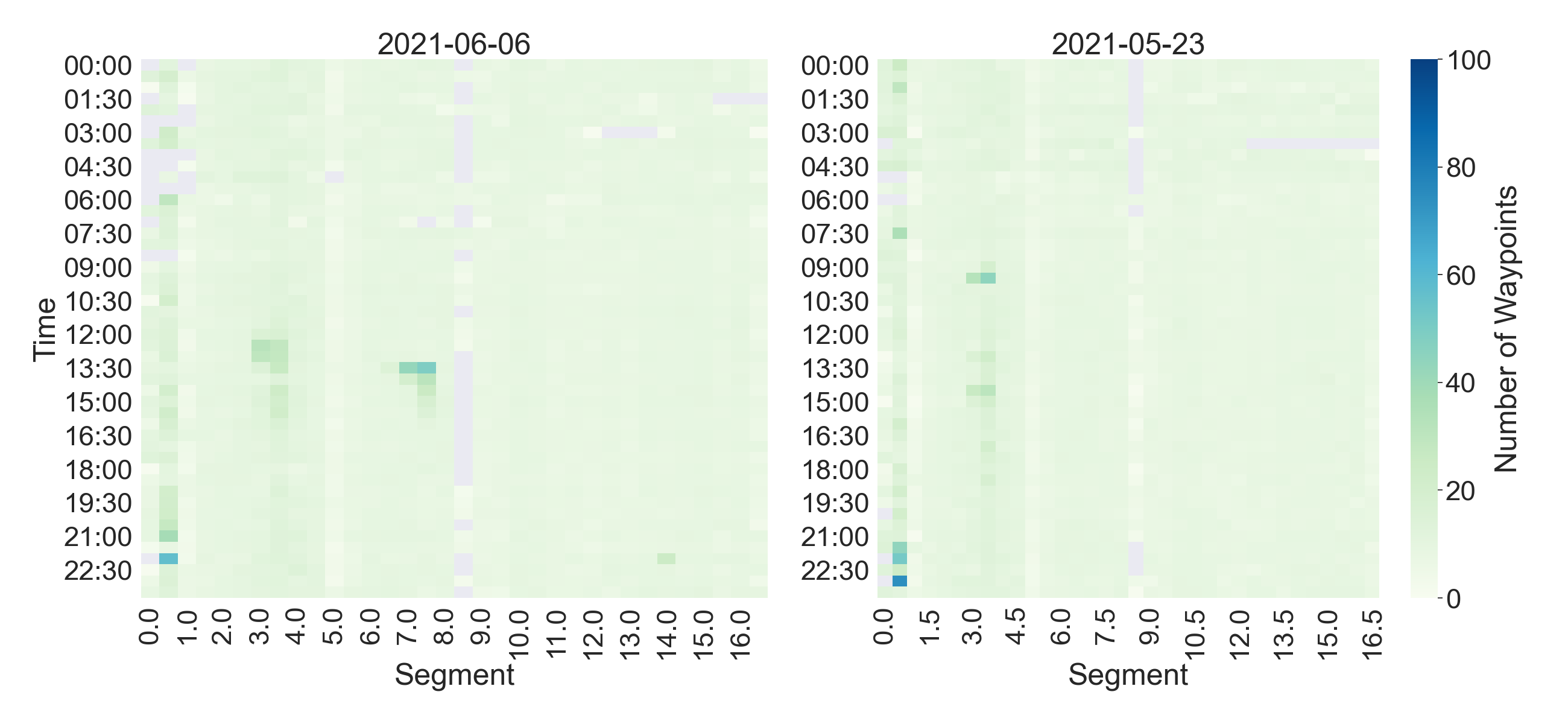}}
\end{minipage}
\caption{Number of waypoints per vehicle }
\label{fig:num_waypoints}
\end{figure}

the influence of fatal crash on both eastbound and westbound traffic can also be revealed by the number of waypoints generated per vehicle every 30-min interval for each 0.5-mile segment. As shown in Fig. \ref{fig:num_waypoints}, the average number of waypoints a vehicle generated along a 0.5-mile segment (either westbound or eastbound) within the 30-min data collection interval is 10 during these two day period. The higher number of waypoints in the weaving area and near toll plaza are expected due to reduced travel speed. Especially, extremely high number of waypoints in eastbound direction (when fatal crash took place) is another sign of major traffic disruption.


\subsection{Profiling metrics}
In this section, we discuss profiling metrics through a combination of different measurements calculated from individual vehicle movement data.
\subsubsection{Safety} as described in the monitoring framework, the safety index is calculated by combining speed coefficient of variation, speed drop percentage from roadway speed limit, and average absolute heading changes over each 0.5-mile segment. Since these measurements capture traffic vulnerability, the higher the index, the unsafe the traffic. In this case study, we set all the coefficients as 1, i.e., all the three measurements weight the same in the overall safety index.

\begin{figure}[h]
\begin{minipage}[h]{0.45\textwidth}
\centering
\subfloat[eastbound]{\includegraphics[scale=0.14]{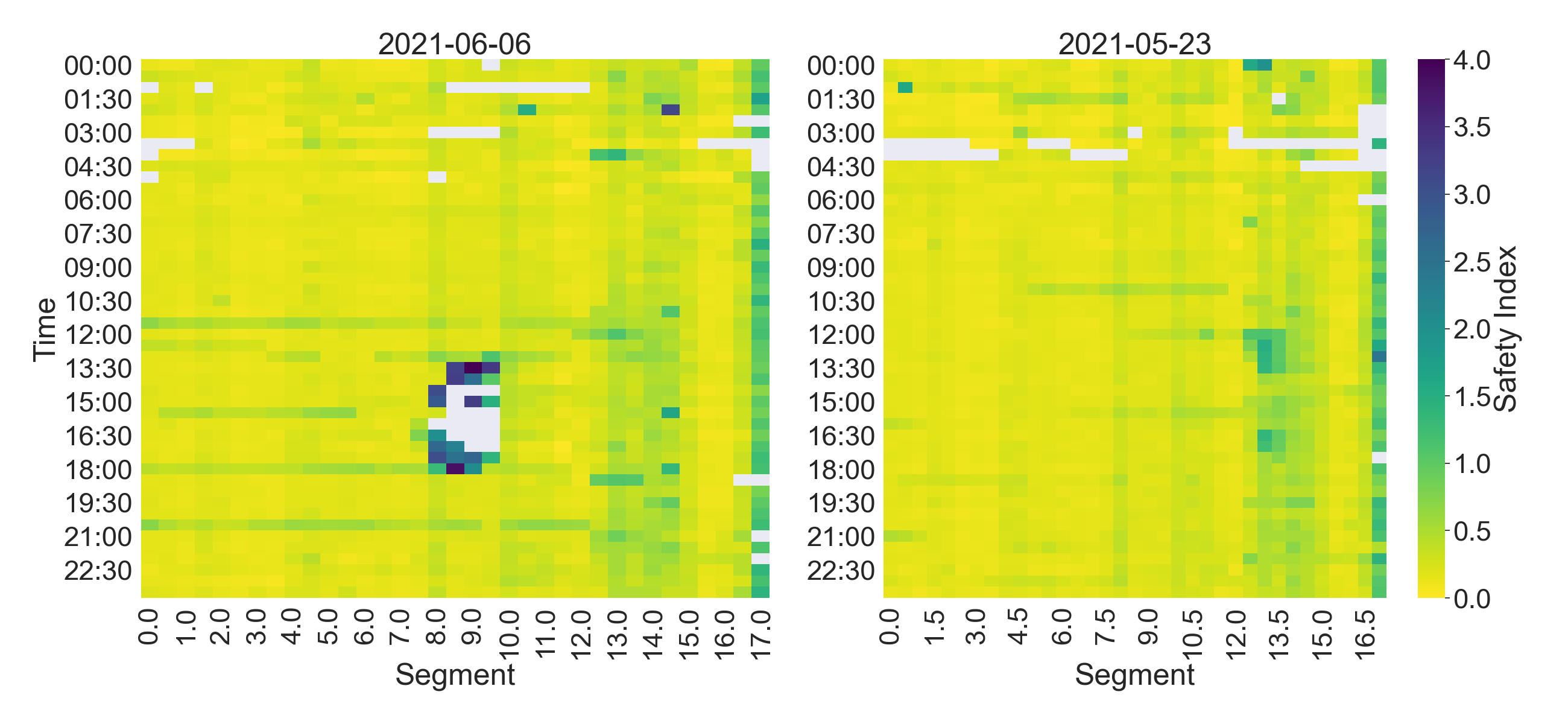}}
\end{minipage} \\
\begin{minipage}[h]{.45\textwidth}
\centering
\subfloat[westbound]{\includegraphics[scale=0.14]{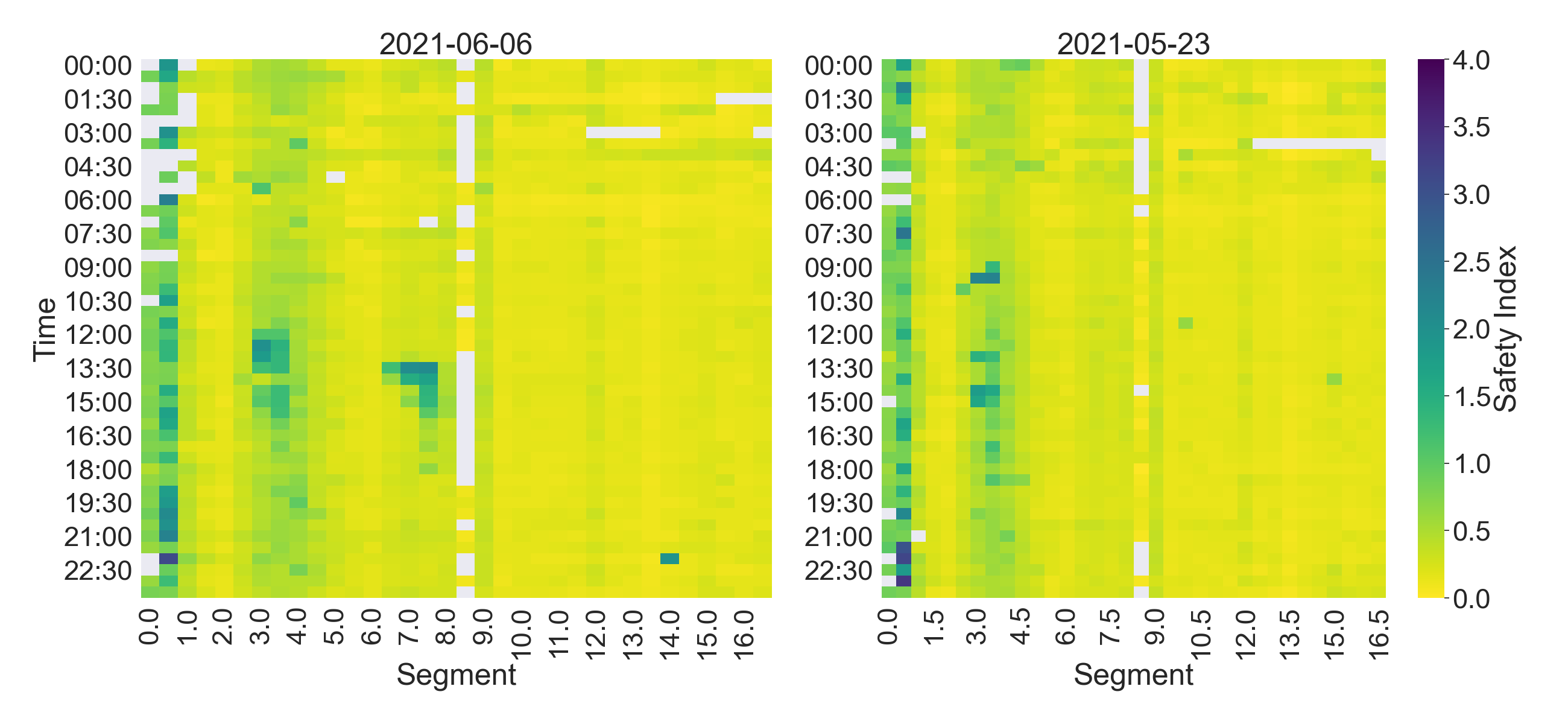}}
\end{minipage}
\caption{Safety index }
\label{fig:safety}
\end{figure}

Fig. \ref{fig:safety} represents the safety index for eastbound and westbound traffic, respectively. The majority of the index is less than 1, indicating smooth and homogeneous traffic conditions, while a higher index means that traffic disruption exists, reflected by large speed variation, speed reduction, or abrupt heading changes over a short distance. Thus, we observe a high score towards the toll plaza at the east end of the roadway, in the area featuring dense ramps, and in the spatial-temporal space with severe crashes.

\begin{figure}[H]
\begin{minipage}[h]{0.45\textwidth}
\centering
\subfloat[eastbound]{\includegraphics[scale=0.14 ]{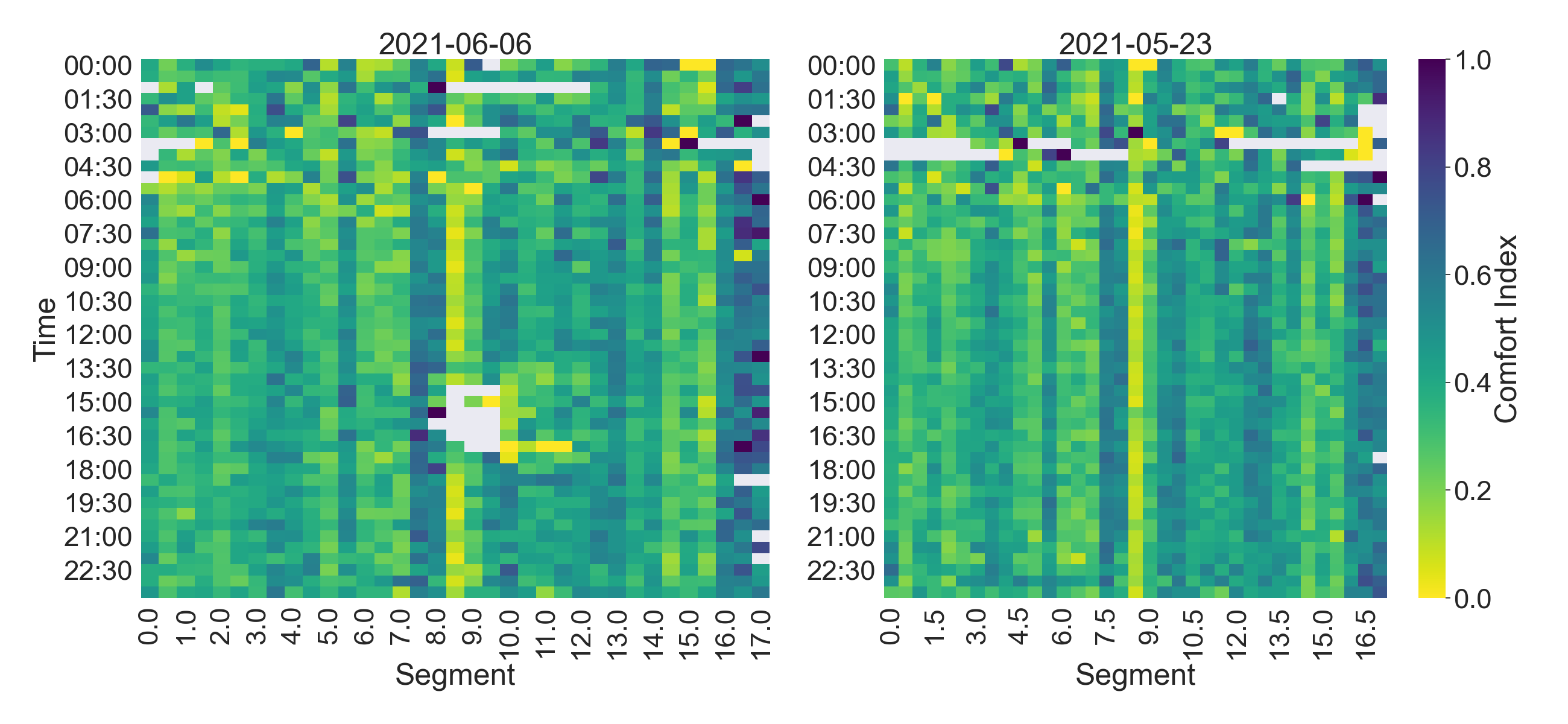}}
\end{minipage} \\
\begin{minipage}[h]{.45\textwidth}
\centering
\subfloat[westbound]{\includegraphics[scale=0.14 ]{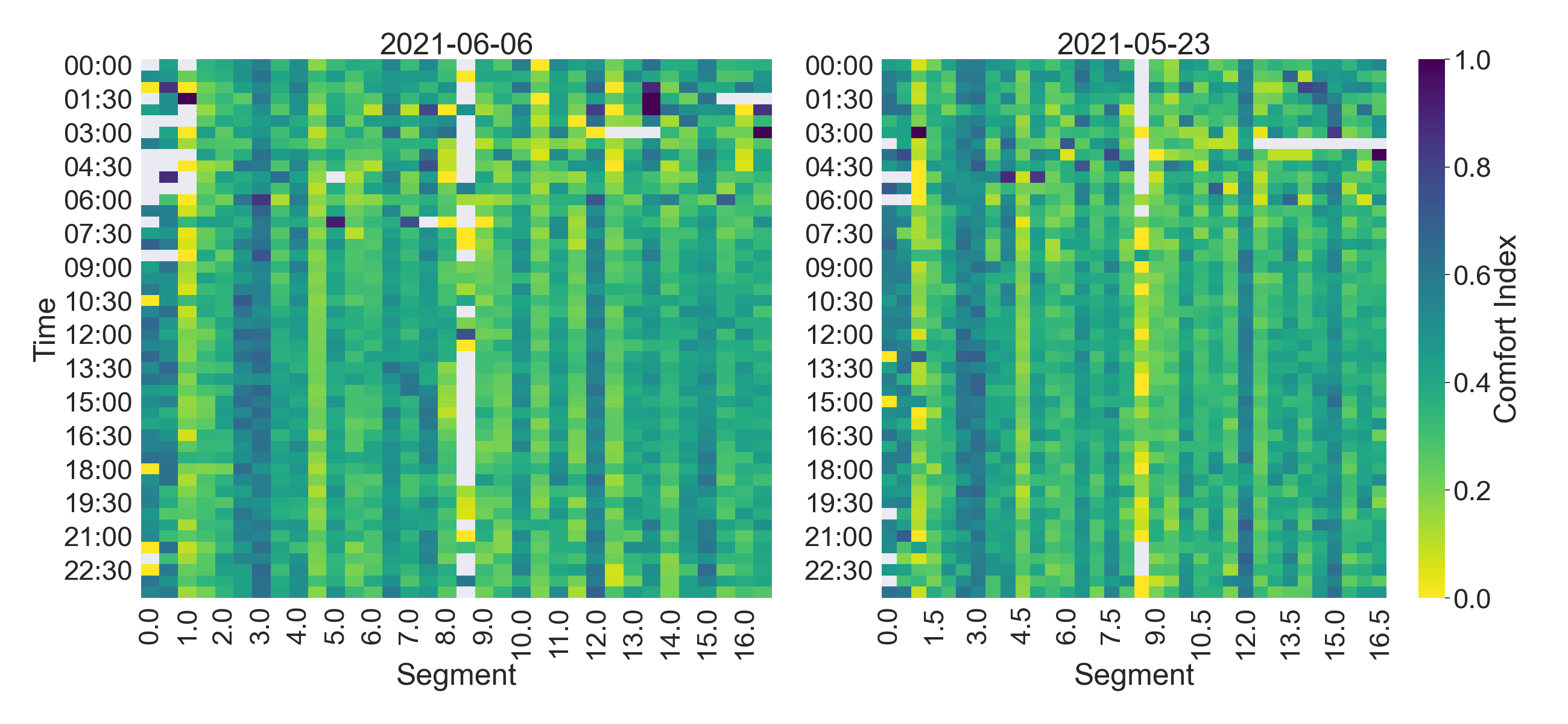}}
\end{minipage}
\caption{Comfort index }
\label{fig:comfort}
\end{figure}

\subsubsection{Comfort}
Similarly, the comfort index is captured by the percentage of brakes and percentage of high jerks: with lower score, few brakes and less abrupt changes of acceleration/braking are observed when vehicles travel through a segment. In Fig. \ref{fig:comfort}, we observe more spatio-temporal grids with higher score in westbound direction. Generally, segments near ramps are featured with higher scores, where more braking and changes of acceleration behavior are involved due to vehicle merging and diverging activities. For instance, milepost 7.5 to 8.5 in eastbound direction, involving two interchanges \cite{njsld}, constantly score high across the day. In contrast, milepost 8.5 to 9.5, without any access points, has relatively low score across the day.

\begin{figure}[h]
\begin{minipage}[h]{0.45\textwidth}
\centering
\subfloat[eastbound]{\includegraphics[scale=0.13]{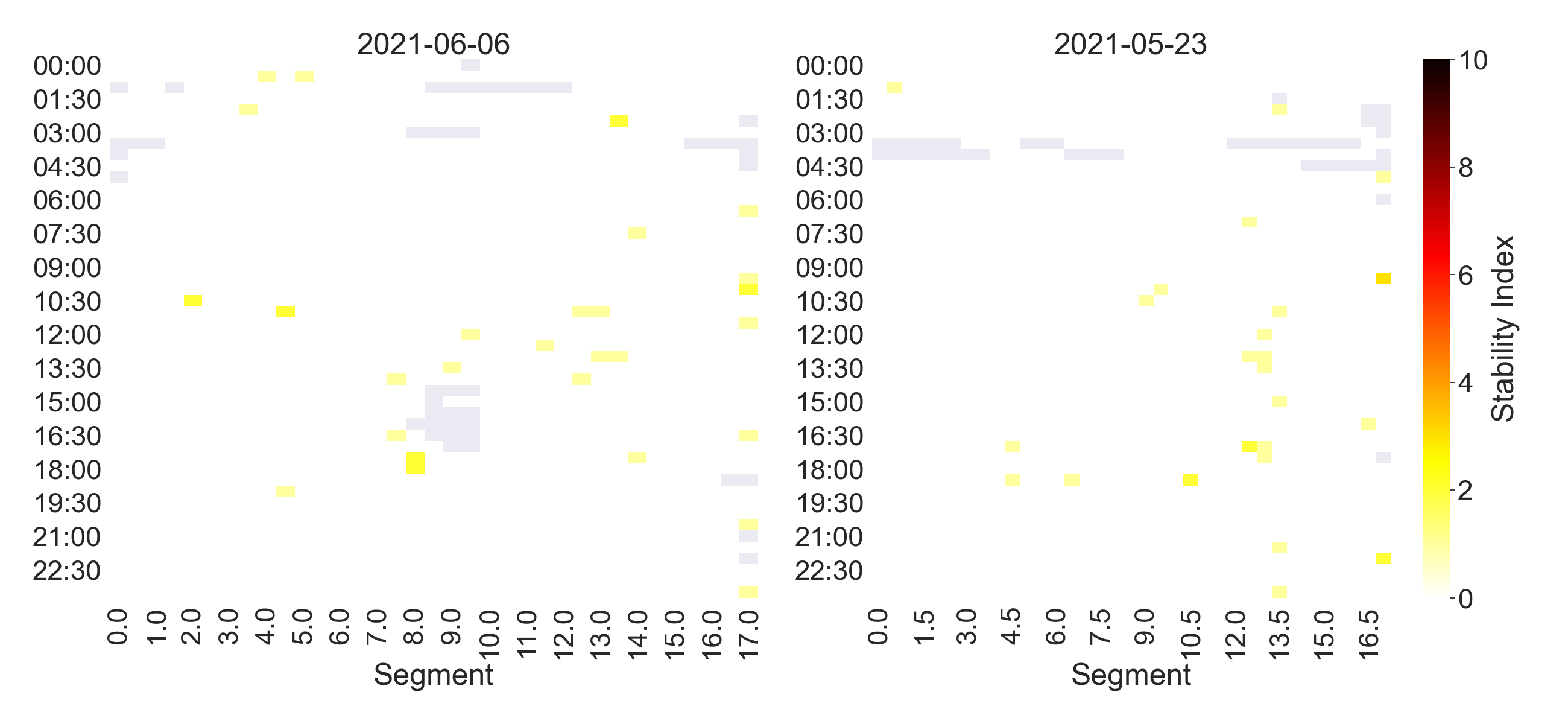}}
\end{minipage} \\
\begin{minipage}[h]{.45\textwidth}
\centering
\subfloat[westbound]{\includegraphics[scale=0.13]{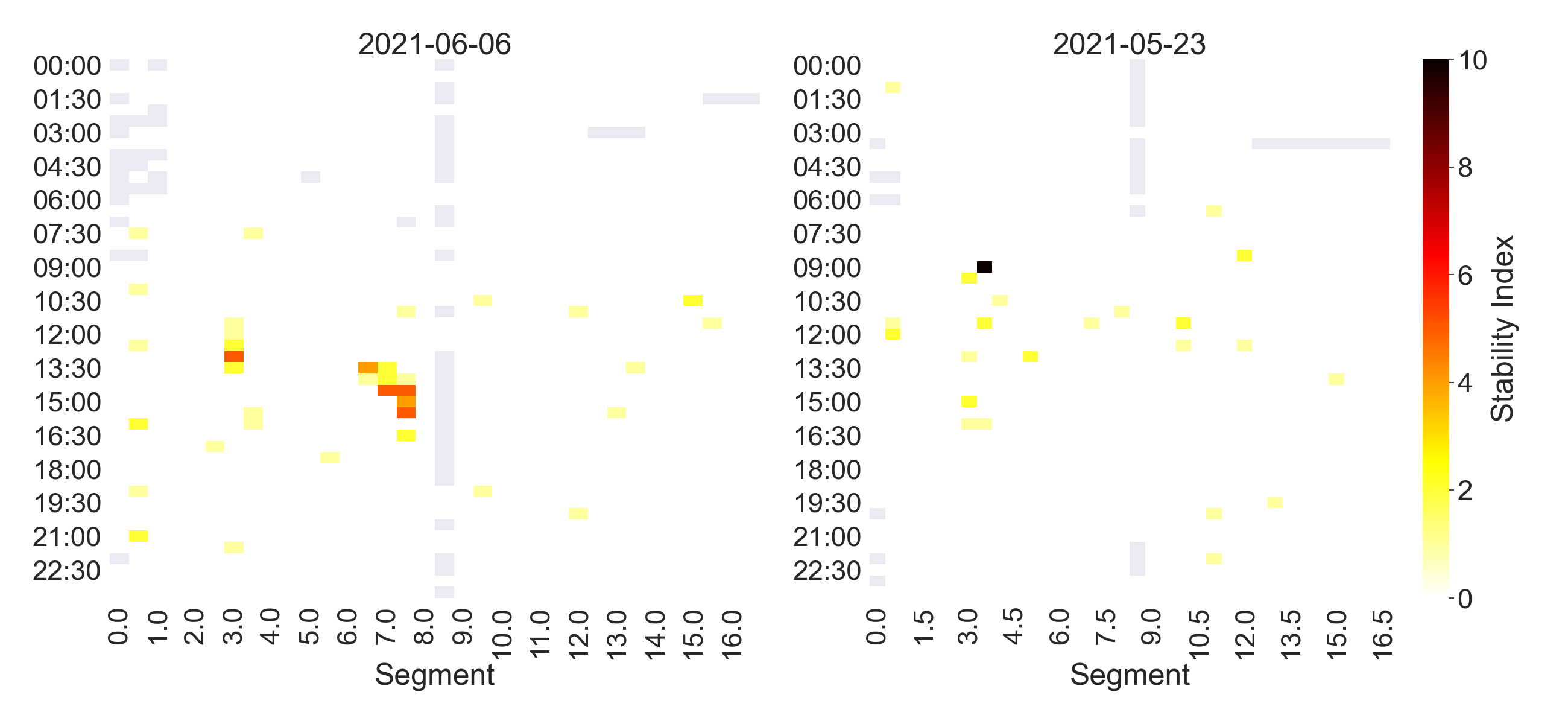}}
\end{minipage}
\caption{Stability index }
\label{fig:stability}
\end{figure}


\subsubsection{Stability}
In this case study, we use the thresholds for hard acceleration ($+3.8m/s^2$) and hard braking ($-2.638m/s^2$) for the definition of large brake/acceleration. The weights for both measures are set as 1 to calculate stability index. Same as the above two indices, the higher score, the unstable the traffic. We see more high value grids in the westbound direction, especially, the eastbound fatal crash leads to a high number of hard acceleration/braking in the corresponding westbound segment. Eastbound milepost 13 to 14 consists of two major exits in Newark urban area, where a lot of weaving activities happens. Thus, we observe frequent high scores, as indicated in Fig. \ref{fig:stability}, the corresponding westbound segment (near milepost 3.0 to 4.0) also features high scores.


\subsubsection{Fuel Consumption}

\begin{figure}[h]
\begin{minipage}[h]{0.45\textwidth}
\centering
\subfloat[eastbound]{\includegraphics[scale=0.14]{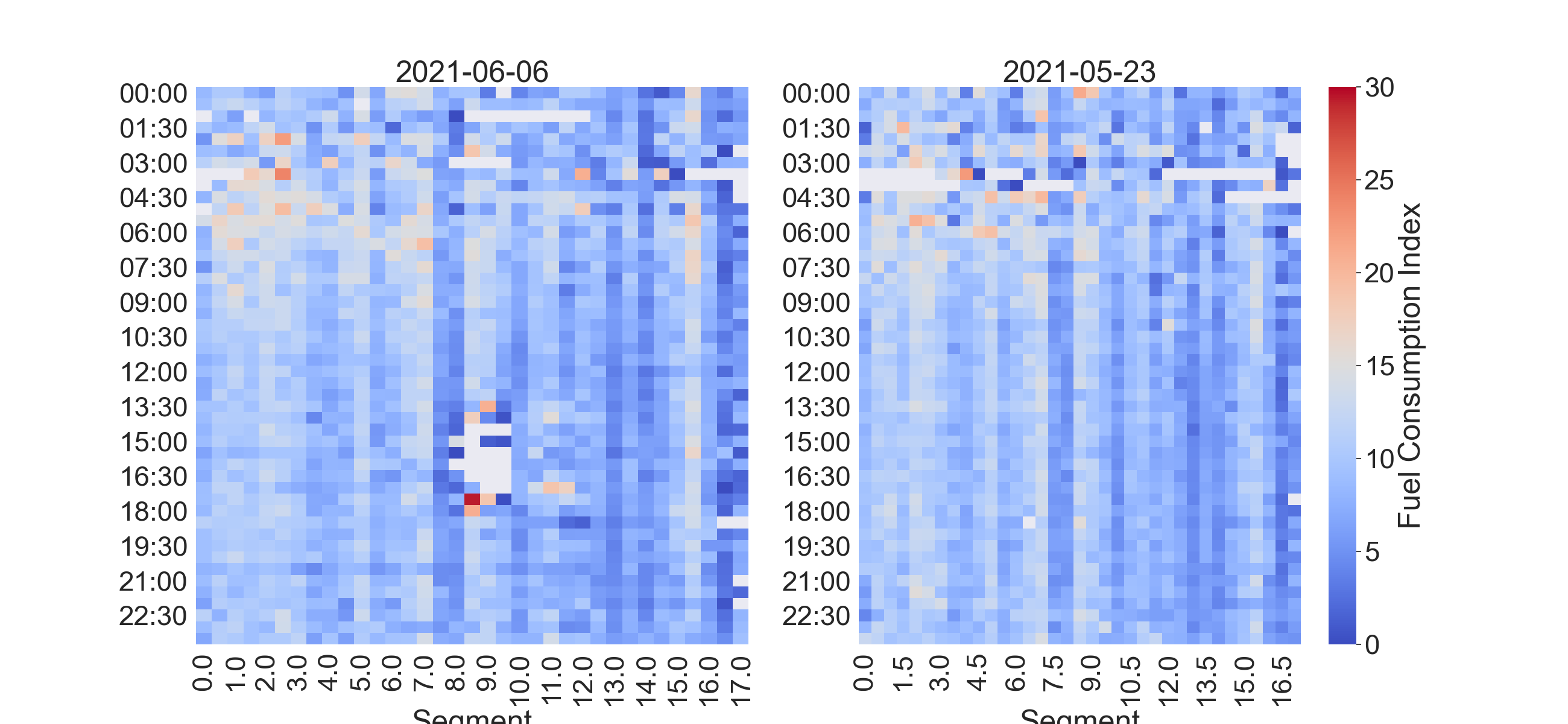}}
\end{minipage} \\
\begin{minipage}[h]{.45\textwidth}
\centering
\subfloat[westbound]{\includegraphics[scale=0.14]{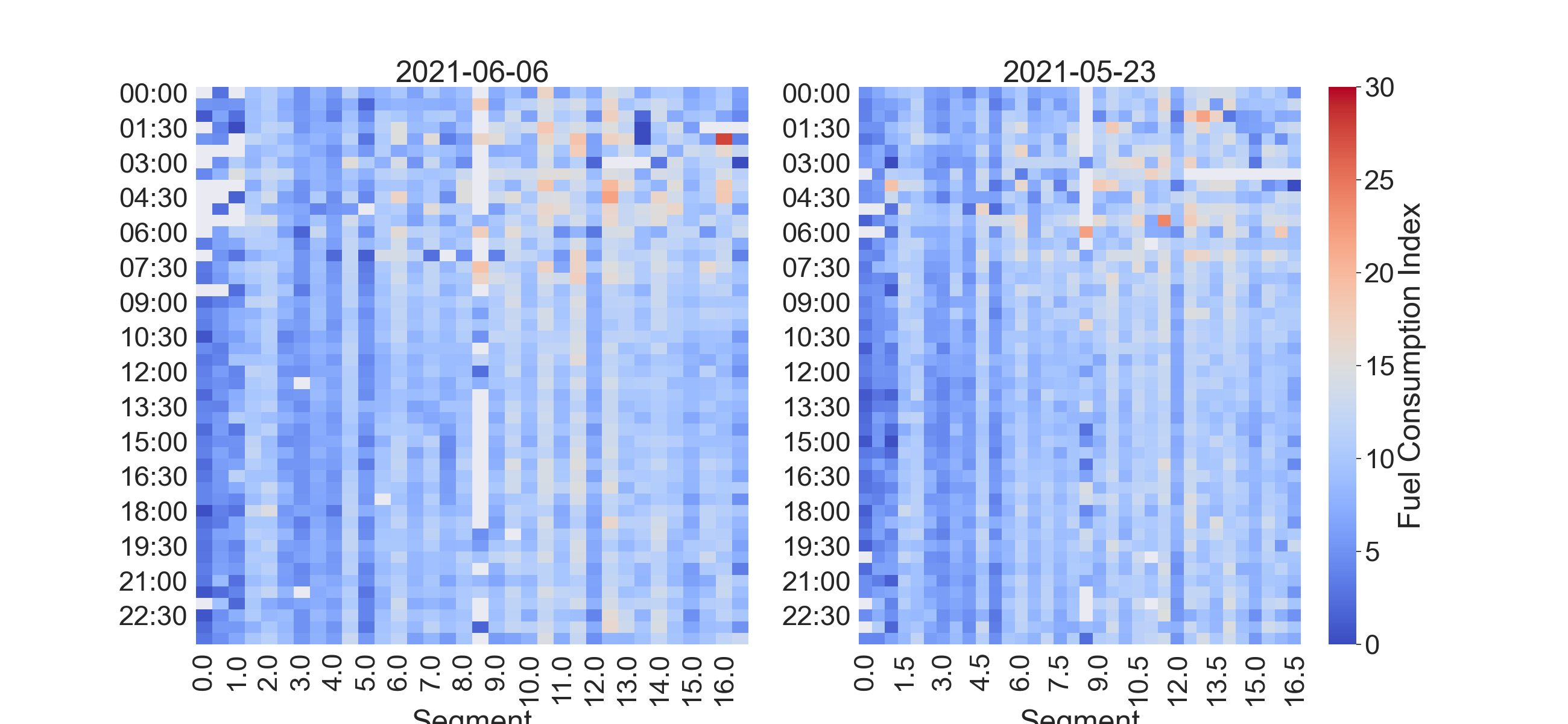}}
\end{minipage}
\caption{Fuel consumption index }
\label{fig:fuel}
\end{figure}


The fuel consumption scan was shown in Fig. \ref{fig:fuel} with each cell represents the average fuel usage for each vehicle. Relatively large fuel consumption were observed in the nighttime, early morning hours. This could be attribute to vehicle driving fast due to the less traffic density during these hours. For the westbound direction, there is a 6\% grade incline between mile markers 8 and 10, this could pose a challenge for certain vehicles, especially heavy-duty ones. Constant higher fuel consumption can be observed during that segment.
The role of a roadway geometry conditions play in the overall traffic profile would be one of the subsequent studies.

Overall, the indices defined here are able to reflect traffic disruptions along the roadway, either from major event or due to complicated roadway geometry. Although the metrics we calculated in the case study are simplified for illustration purpose with all weights set as 1, different values can be set in practice for a variety of applications. For real-time traffic monitoring purpose, a baseline could be established with different thresholds set such that any abnormal traffic conditions and potential disruptions could be identified for proactive traffic management purpose.

\section{Conclusion}
In this paper, we investigated the application of utilizing high resolution vehicle movement data for highway traffic profiling, with different indices calculated from the raw data. With all available trajectory details, it is possible to go beyond the traditional traffic metrics, such as average speed, traffic volume and density. We briefly discussed potential indices that could be obtained from the data, including safety index involving vehicle speed and heading information, comfort index involving vehicle acceleration and jerk information, stability index calculated from vehicle acceleration and braking activities and fuel consumption based on vehicle speed and acceleration. Thus, a whole picture of traffic conditions is obtained over any given roadway segment during a pre-defined time interval (or in a real-time manner), which eventually enables active traffic monitoring and efficient traffic management practices.


\bibliographystyle{IEEEtran}
\bibliography{Wejo-Risk}
\end{document}